\RequirePackage{fix-cm}
\documentclass[twocolumn]{svjour3}          
\smartqed  
\usepackage{graphicx,amssymb}
\begin{document}

\title{Magnetized Plasma Target for Plasma-Jet-Driven Magneto-Inertial Fusion}\thanks{This work 
was supported by the U.S. Department of Energy under contract no.\ DE-AC52-06NA25396.}


\author{Scott C. Hsu \and Samuel J. Langendorf}

\institute{Scott C. Hsu \at
	Los Alamos National Laboratory, Los Alamos, NM 87545 \\
	Tel.: +1-505-667-3386\\
	\email{scotthsu@lanl.gov} 
	\and
	Samuel J. Langendorf \at
	Los Alamos National Laboratory, Los Alamos, NM 87545 \\
	Tel.: +1-505-667-9292\\
	\email{samuel.langendorf@lanl.gov} 
}         
              

\maketitle

\begin{abstract}
We identify the desired characteristics and parameters of a $\beta> 1$
magnetized plasma, possibly with highly tangled, open field lines,
that could be a suitable target to be compressed to fusion conditions
by a spherically imploding plasma liner [S. C. Hsu et al., IEEE Trans.\ Plasma Sci.~{\bf 40}, 1287 
(2012)] formed by merging hypersonic plasma jets.  This 
concept is known as plasma-jet-driven magneto-inertial fusion (PJMIF)\@.
We set requirements on the target and liner such that (a)~compressional
heating dominates over thermal transport in the target, and
(b)~magnetic amplification due to compression dominates over dissipation
over the entire implosion.  We also evaluate the requirements to avoid
drift-instability-induced anomalous transport and current-driven anomalous resistivity in the
target.  Next, we describe possible approaches to create such a magnetized, $\beta>1$
plasma target.  Finally, assuming such a target can be created,
we evaluate the feasibility of a proof-of-concept
experiment using presently achievable plasma jets to demonstrate target compressional heating
at a plasma-liner kinetic energy of $\lesssim 100$~kJ (a few hundred times below that
needed in a PJMIF reactor).
\end{abstract}

\section{Introduction}
\label{sec:intro}

This paper identifies the desired characteristics and parameters of a magnetized plasma target
for plasma-jet-driven magneto-inertial fusion (PJMIF) \cite{thio99,hsu12ieee,knapp14,langendorf17pop}, 
which is a reactor-friendly magneto-inertial-fusion (MIF, aka magnetized target fusion)
\cite{lindemuth83,kirkpatrick95,drake96ft,lindemuth09,wurden16} concept.
PJMIF aims to employ a spherically imploding plasma liner, formed by merging an array of 
hypersonic plasma jets, as a standoff driver to repetitively compress
a magnetized target plasma to fusion conditions.

The PJMIF concept was invented in the late 1990s \cite{thio99}.  PJMIF research began with 
identifying the parameter space for the plasma liners and plasma jets required to reach
fusion conditions \cite{thio99,knapp}, and developing the key concepts needed for plasma guns to 
produce the required jet parameters \cite{thio02,cassibry04thesis,cassibry06,thio07}.
These concepts include (1)~electrode contouring, (2)~formation of a pre-ionized
plasma slab prior to acceleration, and (3)~acceleration of the plasma slab in
a non-snowplow mode.
In 2004, HyperV Technologies Corp.\ embarked on designing, fabricating,
and testing contoured coaxial plasma guns \cite{witherspoon09} to validate these concepts.
Plasma-gun development is now continuing with HyperJet Fusion Corporation.
Preparation for a demonstration of plasma-liner formation via merging hypersonic
plasma jets began in 2009 \cite{hsu09,hsu12ieee,hsu12pop,merritt13,merritt14}.
A proof-of-concept experiment to demonstrate the latter
and develop an understanding of ram-pressure scaling \cite{awe11,davis12,cassibry13} and
non-uniformity evolution \cite{cassibry12,kim13} during plasma-liner convergence is now underway 
\cite{hsu18ieee,hsu18jfe} 
on the Plasma Liner Experiment (PLX) \cite{hsu15jpp} at Los Alamos National Laboratory.

However, little research has been devoted thus far to develop a compatible magnetized plasma
target that takes advantage of the standoff and high implosion speed ($>50$~km/s)
of a spherically imploding
plasma liner.  Most MIF experimental target-development efforts over the past many decades
have focused on compact toroids such 
as spheromaks \cite{laberge13} and field-reversed configurations (FRCs)
\cite{intrator04,slough11}, which both must contend with the
onerous challenges of global magnetohydrodynamic (MHD) instabilities and anomalous thermal 
transport associated with micro-instabilities in $\beta\le1$ plasmas.  It is worth noting that the 
best success to date in MIF research,
achieved by the Magnetized Liner Inertial Fusion (MagLIF) program \cite{slutz10,cuneo12ieee} at
Sandia National Laboratories,
is based on the compression of a $\beta\gg1$ magnetized plasma target,
which was imploded at about 70~km/s to reach multi-keV temperatures \cite{gomez14}.

\begin{figure}[!b]
\centerline{\includegraphics[width=2.truein]{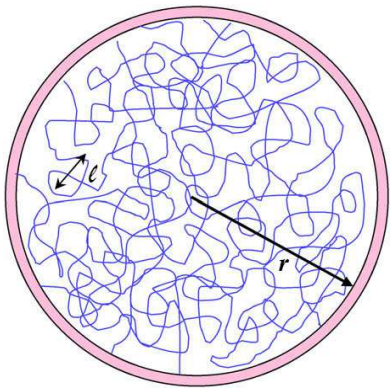}}
\caption{\label{fig:ryutov}Illustration adapted from \cite{ryutov09}
of a hypothetical spherical plasma target with highly tangled, open magnetic field lines
with correlation length $\ell$ $\ll$ radius $r$.}
\end{figure}

The high implosion speed of plasma liners opens up new options for targets,
i.e., $\beta > 1$, ``wall-confined'' plasmas  (prior to compression)
that have Hall magnetization parameters
$\omega_{i}\tau_{i}\gtrsim 1$ and $\omega_e\tau_e \gg 1$ (where $\omega_{i,e}$ are the ion and 
electron cyclotron frequencies,
respectively, and $\tau_{i,e}$ the ion and electron collision times, respectively) to
benefit from magnetized perpendicular thermal transport, while potentially sidestepping 
the issue of magnetohydrodynamic (MHD) instabilities.  The latter statement remains to be explored 
and demonstrated further.  Such a pre-compression, $\beta> 1$, wall-confined plasma
may have closed or highly tangled, open field lines.  The latter is
depicted in Fig.~\ref{fig:ryutov} 
and was previously discussed as a potential MIF target \cite{ryutov09} having
a force-free, ``magneto-static'' tangled magnetic
field \cite{ryutov02ppcf} with initial and instantaneous (during compression) 
correlation lengths $\ell_0$ and $\ell$, respectively.
For self-similar target compression, $\ell_0/r_0 = \ell/r$, where
$r_0$ and $r$ are the initial and instantaneous target
radii during compression, respectively.

For the case of closed field lines, magnetized perpendicular thermal transport dominates and is
controlled by the ions.  For the case of highly tangled, open field lines, 
parallel thermal transport dominates and is
controlled by the electrons.  If the electron mean free path
$\lambda_e$ is sufficiently short, i.e., $\lambda_e \ll \ell$, and 
if the magnetic-field connection length $L\sim r^2/\ell \gg r$ \cite{ryutov09} is very long,
i.e., $\ell/r\ll 1$, then diffusive electron transport along $L$
may allow for a sufficiently long energy-confinement time such that 
heating due to target compression could dominate over thermal losses.  
The interesting situation with $\beta>1$ and highly tangled, open field lines may also exist in 
astrophysical molecular clouds \cite{ryutov05}.

This paper is organized as follows.  In the ``Fusion-Scale
PJMIF Target'' section, we describe
the requirements of a fusion-scale magnetized plasma target such that (a)~compressional 
heating and magnetic amplification dominate over thermal losses and magnetic dissipation,
respectively,
and (b)~anomalous thermal transport associated with drift instabilities
and anomalous resistivity associated with current-driven instabilities
are avoided.
In the ``Possible Approaches to $\beta> 1$ Target Formation'' section, we describe 
conceptual ideas on how such
a $\beta >1$, PJMIF-compatible target might be formed, with the goal of guiding future
research efforts in this area.
In the ``Feasibility of a Near-Term Target-Heating Experiment'' section,
we evaluate the feasibility of a subscale experiment
to demonstrate target heating via compression by a spherically impoding plasma liner, with
kinetic energy $\lesssim 100$~kJ, that can be formed by the existing generation of coaxial plasma 
guns \cite{witherspoon17,hsu18ieee}.  The paper ends with ``Conclusions and Future Work.''

\section{Fusion-Scale PJMIF Target}
\label{sec:PJMIF_target}

In this section, we identify the properties of a $\beta > 1$, $\omega_i\tau_i\gtrsim 1$
magnetized target plasma that is ideally
suited for the high implosion speed of spherically imploding plasma liners \cite{hsu12ieee}.
The {\em pre-compression} target parameters given in Table~\ref{table:PJMIF_target},
\begin{table}[!t]
\caption{\label{table:PJMIF_target} Nominal, fusion-scale, {\em pre-compression} target parameters,
assuming ion-to-proton mass ratio $\mu=2.5$ (DT)\@.}
\begin{center}
\begin{tabular}{ll}
\hline\noalign{\smallskip}
parameter & value\\
\noalign{\smallskip}\hline\noalign{\smallskip}
radius $r_0$ & 4~cm\\
density $n_0$ & $1 \times 10^{18}$~cm$^{-3}$\\
temperature $T_{e0}=T_{i0}$ & 80~eV\\
pressure $p_0$ & 25.6~MPa\\
magnetic field $B_0$ & 4.5~T\\
thermal energy $E_0$ & 10.3~kJ\\
\hline
thermal/magnetic pressure ratio $\beta$ & 3.2\\
ion Hall parameter $\omega_i\tau_i$ & 0.5\\
electron Hall parameter $\omega_e\tau_e$ & 25\\
ion mean free path $\lambda_i$ & 0.015~cm\\
electron mean free path $\lambda_e$ & 0.012~cm\\
ion gyro-radius $\rho_i$ & 0.032~cm\\
electron gyro-radius $\rho_e$ & 0.00047~cm\\
tangled-field scale $\ell_0$ & 0.4~cm\\
implosion speed $v_0$ & 100~km/s\\
\noalign{\smallskip}\hline
\end{tabular}
\end{center}
\end{table}
which are modified from the target parameters resulting in 1D calculated
energy gain greater than unity upon compression by a plasma liner \cite{langendorf17pop},
has $\beta > 1$ and $\omega_i\tau_i\sim1$.
In particular, we are interested in its
macro-stability, thermal-transport, and magnetic-dissipation properties during
compression, as these properties directly impact its utility as an MIF target plasma.

\begin{figure}[!b]
\centerline{\includegraphics[width=2.3truein]{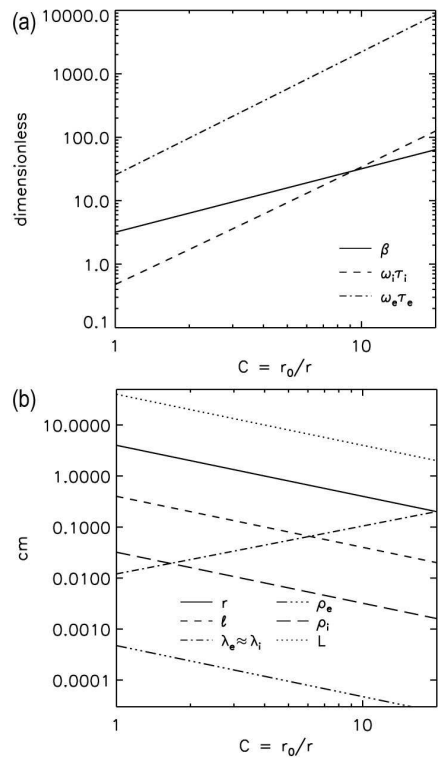}}
\caption{\label{fig:target_vs_C} Quantities in the legends
vs.\  $C$ for the fusion-scale, pre-compression target parameters of
Table~\ref{table:PJMIF_target}, assuming adiabatic compression.}
\end{figure}

For adiabatic spherical compression (i.e., assuming $pV^\gamma =$~constant, where
$p$ is the thermal pressure,
$V$ is the plasma volume, and polytropic index $\gamma = 5/3$), the scalings of target parameters
with convergence ratio $C\equiv r_0/r$ are
$n=n_0 C^3$, $T_e=T_{e0} C^2$, $T_i = T_{i0} C^2$, $B=B_0 C^2$, 
thermal energy $E=E_0 C^5$, and $\beta\sim \beta_0 C$\@.
Based on these relations and the parameters in 
Table~\ref{table:PJMIF_target},
Fig.~\ref{fig:target_vs_C}(a) shows that for
the entire implosion (or very nearly so), $\omega_i\tau_i > 1$,
$\omega_e\tau_e\gg 1$, and $\beta\gg 1$.  Figure~\ref{fig:target_vs_C}(b)
shows that for the entire implosion up to $C\gtrsim 10$,
$\rho_i, \rho_e\ll r$, $\lambda_i \approx \lambda_e \ll r$,
and $\lambda_e/L\ll 1$.  Thus, it is possible that this target avoids MHD instabilities
(because $\beta\gg 1$), and exhibits near-classical perpendicular (because 
$\omega_i\tau_i, \omega_e\tau_e >1$ and $\rho_i,\lambda_i \ll r$) and parallel 
(because $\rho_e\ll \ell$ and $\lambda_e\ll L$) thermal transport.  These properties
are examined in more detail next.

\subsection{Macro-stability of a Fusion-Scale Target}
\label{sec:PJMIF_stability}

One potential advantage of a $\beta\gg1$ target plasma is that it may sidestep the
issue of fast, virulent MHD instabilities that plague a $\beta < 1$ target
before it can even get compressed by a liner.
In a $\beta\gg1$ target, the magnetic pressure is and remains very small through
stagnation (because $\beta\sim C$ for adiabatic spherical convergence)
compared to the thermal pressure, and therefore 
hydrodynamic instabilities (e.g., Rayleigh-Taylor or Kelvin-Helmholtz)
at the liner/target interface become the primary stability concern,
particularly during deceleration just prior to stagnation.  This is an important issue,
beyond the scope of this paper, requiring
much further study in the context of PJMIF\@.
The hydrodynamic disassembly time
($\sim 0.5$~$\mu$s for the hypothetical target in Table~\ref{table:PJMIF_target}) becomes a bottleneck in that the incoming liner must engage and start compressing
the target before it can expand very much (also an issue requiring further study).
It must also compress the target in a short-enough time such that compressional 
heating and magnetic amplification overcome the thermal loss rate and magnetic dissipation,
respectively, during target compression.

\subsection{Thermal Transport in a Fusion-Scale Target}
\label{sec:thermal}

In this subsection, we evaluate the requirement for target compressional heating
to dominate over thermal loss rates, which have
characteristic time $\tau_E$.  The evolution of the target thermal energy $E$
during compression (ignoring radiative losses, justified below) is
\begin{equation}
\frac{{\rm d}E}{{\rm d}t} = \left(-\frac{5}{r}\frac{{\rm d} r}{{\rm d}t} - \frac{1}{\tau_E}\right) E,
\label{eq:E}
\end{equation}
where the first term on the right hand side is compressional heating power
${\rm d}(E_0 r_0^5/r^5)/{\rm d}t =-(5E/r)({\rm d}r/{\rm d}t)$.
For the latter to dominate over thermal losses, the requirement is
\begin{equation}
\tau_E \gg \frac{r}{5|{\rm d}r/{\rm d}t|} \approx \frac{r}{5v_0}\equiv \tau_{E,req},
\label{eq:adiabatic}
\end{equation}
where we assume that ${\rm d}r/{\rm d}t = -v_0$.
We consider three specific cases governing $\tau_E$ and require 
Eq.~(\ref{eq:adiabatic}) to be satisfied in each case:
\begin{enumerate}
\item For a target with closed field lines, the
classical perpendicular ion diffusion time $\tau_{Ei,\perp}\sim r^2/D_i \gg \tau_{E,req}$
is required, where 
$D_i= \rho_i^2 \nu_i$ is the perpendicular ion diffusivity.
\item For a target with closed field lines, the perpendicular
anomalous diffusion time $\tau_B\sim r^2/D_B\gg \tau_{E,req}$ is also required,
where $D_B =kT/16eB$ is the Bohm diffusivity; if anomalous transport does not arise (discussed 
below), then this condition is not a requirement.
\item For a target with highly tangled, open field lines,
the classical parallel electron diffusion time $\tau_{Ee,\|} {\rm (R)} \sim L^2/D_{e,\|} \gg \tau_{E,req}$ 
is required,
where ``(R)'' stands for Ryutov \cite{ryutov09} 
and $D_{e,\|} = v_{te}^2/\nu_e$ is the parallel electron diffusivity; if, however, adjacent field
lines diverge exponentially \cite{rechester78}, the effective {\em perpendicular} diffusivity may 
become $D_{e,\|}/[3\ln(\ell/\rho_e)]$~\cite{chandran98},
resulting in a faster thermal loss time
$\tau_{Ee} {\rm (CC)} \sim 3\ln(\ell/\rho_e)(r^2/D_{e,\|})$, where ``(CC)'' stands for Chandran and 
Cowley \cite{chandran98}.  For the parameters of Table~\ref{table:PJMIF_target},
$\tau_{Ee} {\rm (CC)} \sim 3\tau_{Ee,\|} {\rm (R)} (\ell/r)^2 \ln(\ell/\rho_e) \approx 
\tau_{Ee,\|} {\rm (R)}/5$.  For $\lambda_e > \ell$,
$\tau_{Ee} {\rm (CC)}$ may increase due to mirror-trapping effects \cite{chandran98,albright01}.
See the appendix.
\end{enumerate}

\begin{figure}[!b]
\centerline{\includegraphics[width=2.3truein]{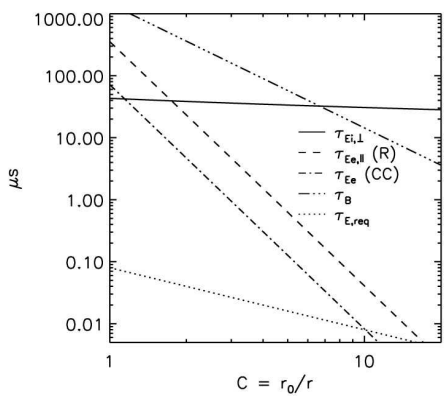}}
\caption{\label{fig:PJMIF_tauE_vs_C} Quantities in the legend vs.\ $C$ for the fusion-scale,
pre-compression target parameters of Table~\ref{table:PJMIF_target}, assuming
adiabatic heating.}
\end{figure}

Using the pre-compression target parameters in Table~\ref{table:PJMIF_target}
and assuming that $\ell_0/r_0 = \ell / r$ \cite{ryutov09}
throughout the target compression, we calculate
$\tau_{Ei,\perp}$, $\tau_{Ee,\|} {\rm (R)}$, $\tau_{Ee} {\rm (CC)}$, $\tau_B$, 
and $\tau_{E,req}$ vs.\ $C$, assuming adiabatic compression,
as shown in Fig.~\ref{fig:PJMIF_tauE_vs_C}.
If $\ell_0$ is too large, then $L$ is too small, resulting
in violation of the required condition that $\tau_{Ee,\|} {\rm (R)} \approx 5\tau_{Ee} {\rm (CC)}
\gg \tau_{E,req}$.  
For a target with closed field lines \cite{ryutov09},
the requirements $\tau_{Ei,\perp},\tau_{B} \gg \tau_{E,req}$
are comfortably satisfied to $C>10$.  For a target with highly tangled, open field lines,
the requirement $\tau_{Ee}\gg \tau_{E,req}$ is challenging to meet at high $C$,
although for the parameters of Table~\ref{table:PJMIF_target},
the requirement is satisfied to $C\approx 10$ for both $\tau_{Ee,\|} {\rm (R)}$ and
$\tau_{Ee} {\rm (CC)}$. A caveat is that we assume
a gradient scale length $r$ for estimating the various diffusion times.  If the gradient
scale length is in fact a fraction of $r$, then the values for $\tau_E$ 
shown in Fig.~\ref{fig:PJMIF_tauE_vs_C} (except $\tau_{E,req}$) will be reduced by the square
of that fraction.  This caveat applies to all the estimates of energy-diffusion and
magnetic-dissipation times in the remainder of the paper.  See the appendix for
a more-detailed treatment on the limits of target adiabatic heating.

Ignoring radiative losses is justified in the previous analysis.  The radiative loss time is
$\tau_R\sim E/P_{Br}$, where the bremsstrahlung power $P_{Br}\sim n_e^2 T_e^{1/2}\sim C^7$ \
\cite{nrl-formulary}, and therefore $\tau_R \sim C^5/C^7 \sim C^{-2}$. 
For radiative losses to be negligible, $\tau_R/\tau_{E,req}\sim C^{-2}/C^{-1}\sim
C^{-1} \gg 1$ is required.  For the parameters of Table~\ref{table:PJMIF_target},
$\tau_R/\tau_{E,req}= 3750$ at $C=1$ and $\tau_R/\tau_{E,req} = 375$ at $C=10$,
and therefore radiative losses are negligible for the entire implosion to $C>10$.

\begin{figure}[!t]
\centerline{\includegraphics[width=2.3truein]{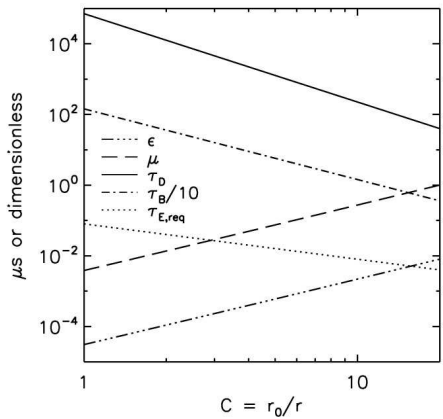}}
\caption{\label{fig:tauD_PJMIF} Quantities in the legend vs.\ $C$ for
the fusion-scale, pre-compression target parameters of Table~\ref{table:PJMIF_target},
assuming adiabatic heating. Even the worse case scenario in a target
with closed field lines, $\tau_B/10$, which is not expected,
meets the requirement $\tau_B/10\gg\tau_{E,req}$.}
\end{figure}

For targets with closed field lines, an important question is whether micro-instabilities
will lead to anomalous perpendicular thermal transport faster than $\tau_B$, i.e.,
is there a mechanism leading to $\tau_E \ll \tau_B$ that becomes the bottleneck
in satisfying Eq.~(\ref{eq:adiabatic})?
The diffusivity associated with drift instabilities in an infinite-$\beta$, collisionless plasma
was shown to be $D \sim 10D_B$ \cite{el-nadi73}.
However, it was later shown \cite{ryutov02pop} that the result would be
different in a high-$\beta$ collisional plasma satisfying the conditions
$\omega_D \ll \nu_i$ and $k_\| \lambda_i \ll 1$, where
$\omega_D$ is the drift frequency and $k_\|$ is the perturbation wave number along $B$\@.
In shearless systems,
the dominant plasma transport arises due to perturbations with the largest possible scale length,
i.e., $k_\| \sim 1/r$.  The two conditions stated above become
$\epsilon\equiv\rho_i \lambda_i/r^2 \ll 1$ and $\mu\equiv\lambda_i/r\ll 1$,
respectively \cite{ryutov02pop}.   When these conditions are satisfied,
$D\sim 4D_B/M_0^{1/2}$ \cite{ryutov02pop}, 
where $M_0=8\mu^2/(\epsilon^2\beta)$, and the drift-instability-induced
diffusion time is $\tau_D\sim r^2/D$.
If the scale size of the magnetic field $\ll r$ (i.e., for a tangled field),
then the analysis of \cite{ryutov02pop} should be reconsidered
because the dominant transport may not be due to perturbations with $k_\| \sim 1/r$.  However, 
$D_B$ and the reduced $D$ remain good benchmarks.
Figure~\ref{fig:tauD_PJMIF} shows that $\epsilon\ll 1$ and $\mu\ll 1$
are satisfied up to $C=10$, which means that perpendicular thermal
transport would be $\tau_D~(\gg \tau_{E,req}$ up to $C>10$).  Also shown
is $\tau_B/10$, which is the nominal worst-case scenario for collisionless, drift-instability-induced
transport.  It can be seen that $\tau_B/10 \gg \tau_{E,req}$, meaning
that drift-instability-induced, anomalous transport should not play a role in this implosion regime.

\subsection{Magnetic-Energy Dissipation in a Fusion-Scale Target}
\label{sec:magnetic}

In this subsection,
we evaluate the requirement that magnetic amplification due to compression is much larger than 
dissipation.  The evolution of magnetic energy density is described by \cite{ryutov09}
\begin{equation}
\frac{\rm d}{{\rm d}t} \left(\frac{B^2}{2\mu_0}\right) = \left(-\frac{4}{r}\frac{{\rm d} r}{{\rm d}t} -
\frac{1}{\tau_M}\right) \frac{B^2}{2\mu_0},
\label{eq:WB_PJMIF}
\end{equation}
where the first term on the right hand side is amplification of magnetic energy
density due to compression, i.e.,
${\rm d}(B_0^2 r_0^4/r^4)/{\rm d}t  = -(4B^2/r)({\rm d}r/{\rm d}t)$,
and $\tau_M$ is the magnetic-diffusion time over length scale $\ell$ of a tangled field
or $r$ of a closed field.  
If magnetic-energy dissipation is negligible, this requires that
\begin{equation}
\tau_M \gg \frac{r}{4 |{\rm d}r/{\rm d}t|}\approx \frac{r}{4v_0} \equiv \tau_{M,req},
\label{eq:B_diss_PJMIF}
\end{equation}
where $\tau_M = \ell^2/D_M$ (we evaluate only the more
demanding case of the tangled field), $D_M=\eta/\mu_0$ is the classical magnetic diffusivity,
and $\eta$ the classical perpendicular resistivity.

\begin{figure}[!t]
\centerline{\includegraphics[width=2.3truein]{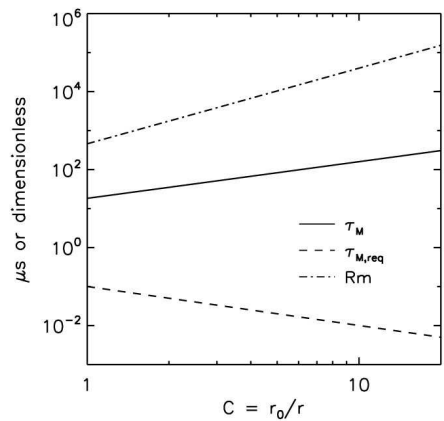}}
\caption{\label{fig:tauM_PJMIF}Quantities in the legend vs.\ $C$ for
the fusion-scale, pre-compression target parameters of Table~\ref{table:PJMIF_target},
assuming adiabatic heating.}
\end{figure}

Figure~\ref{fig:tauM_PJMIF} shows that Eq.~(\ref{eq:B_diss_PJMIF}) is well satisfied
for all $C$ and is more restrictive (although not very) early in the implosion
because $T_e$ is lower, and $D_M$ and $\ell$ are higher.
Figure~\ref{fig:tauM_PJMIF} also shows that the magnetic Reynolds number $Rm\equiv
\mu_0 \ell v_0/\eta \gg 10^2$ (using the more stringent scale $\ell$) for the entire target convergence.
This means that the magnetic field is frozen into the plasma motion,
and that any tangled field that is initially present in the target could indeed be
compressed self-similarly as assumed in \cite{ryutov09}.

We also evaluate the Nernst flux loss, which arises due to finite $\nabla T_e \times  \mathbf{B}$
and can be expressed as \cite{braginskii65,mcbride15}
\begin{equation}
\frac{{\rm d}\Phi}{{\rm d}t} = -\beta_\wedge 2\pi r |\nabla T_e |,
\end{equation}
where
\begin{equation}
\beta_\wedge=\frac{1.5(\omega_e\tau_e)^3 + 3.053\omega_e\tau_e}
{(\omega_e\tau_e)^4 + 14.79(\omega_e\tau_e)^2 + 3.7703}
\end{equation}
is the dimensionless Braginskii thermoelectric coefficient \cite{braginskii65},
$2\pi r$ the circumference of the plasma out of which field
is being advected, and $\nabla T_e$ given in units of V/m.  For adiabatic
spherical convergence, ${\rm d}\Phi/ {\rm d}t$ is independent of $C$, and thus if the Nernst
flux loss is negligible initially, it will be negligible for all $C$.  For the parameters in 
Table~\ref{table:PJMIF_target}, $\Phi\approx B\pi r^2/2 = 
22.6$~mWb, $\beta_\wedge = 0.058$, $\nabla T_e \sim T_e/r =2000$~V/m, 
${\rm d}\Phi/{\rm d}t = 29.08$~Wb/s, and the characteristic Nernst flux-loss time is
\begin{equation}
\tau_N \sim \frac{\Phi}{{\rm d}\Phi/{\rm d}t}  \approx 
\frac{22.62~{\rm mWb}}{29.08~{\rm Wb/s}} \approx 778~\mu{\rm s},
\label{eq:tauN}
\end{equation}
which is $\gg \tau_{M,req}$ and therefore negligible for the fusion-scale target of 
Table~\ref{table:PJMIF_target}.

Finally, we consider
the constraint provided by the condition to avoid current-driven
anomalous resistivity \cite{davidson75},
\begin{equation}
u = \frac{j}{en} \sim \frac{B}{\mu_0 \ell en} < v_{ti},
\label{eq:u_lt_vti}
\end{equation}
where $u$ is the relative drift speed between electrons and ions, $j$
the current density, $e$ the electron charge, and $v_{ti}$ the ion thermal speed.
This condition is also most restrictive
at $r=r_0$, for which Eq.~(\ref{eq:u_lt_vti}) can be rewritten as \cite{ryutov02pop}
\begin{equation}
\left(\frac{r_0}{\ell_0}\right)^2 < \frac{\beta_0}{2}\left(\frac{\omega_{pi0}r_0}{c}\right)^2,
\label{eq:a_over_ell0}
\end{equation}
where $\omega_{pi0}$ is the pre-compression ion plasma frequency.
Using the parameters in Table~\ref{table:PJMIF_target}, we obtain
the requirement that $\ell_0 > 0.3$~mm, which is satisfied for the choice 
$\ell_0=4$~mm.

Thus, for the pre-compression, fusion-scale DT
target parameters of Table~\ref{table:PJMIF_target}, near-adiabatic heating and magnetic flux 
compression with very small dissipation are theoretically possible up to $C\approx 10$.

\section{Possible Approaches to $\beta>1$ Target Formation}
\label{sec:target_formation}

As mentioned in the ``Introduction'' section, formation of a $\beta>1$, wall-confined magnetized
target plasma was previously discussed by D.~Ryutov \cite{ryutov09}, in which he 
states: ``Creation of the initial plasma with a small-scale, random, $\beta\sim 1$ magnetic field immersed into it may not be a simple
task.  The author is not aware of any published papers where formation and characterization
of such an object would be documented.  An intuitively appealing way for creating such a target would
be the use of numerous plasma guns generating small-scale, magnetized plasma bunches and
injection of such bunches into a limited volume.  This could be a version of the guns envisioned
in the plasma liner approach.''  In essence, our primary
target-formation development path has been identified.  We are also interested in
forming a $\beta>1$ target with closed field lines, which may also be explored
by the methods discussed next.

A key to creating a $\beta > 1$, $\omega_i\tau_i \gtrsim 1$
magnetized plasma via merging multiple gun-formed plasmas 
is potentially via adjustment of the
gun parameter $\lambda_{\rm gun} \equiv \mu_0 I_{\rm gun} / \psi_{\rm gun}$, where
$I_{\rm gun}$ and $\psi_{\rm gun}$ are the gun electrical current
and the pre-applied poloidal magnetic flux (``bias flux'')
linking the two gun electrodes, respectively.  Prior research, e.g., \cite{yamada90,ono99,cothran03},
demonstrated that merging two $\beta \ll 1$ spheromaks (with $\lambda_{\rm gun}$
exceeding some threshold value $\lambda_{\rm spheromak}$ depending on gun geometry) 
results in either a spheromak (co-helicity
merging) or an FRC (counter-helicity merging), both of which have $\beta \le 1$,
which we do not want.
At the opposite extreme, where $\psi_{\rm gun}=0$ and $\lambda_{\rm gun}=\infty$,
as is the case with PLX plasma guns \cite{witherspoon11,witherspoon17}, which produce
$\beta\gg 1$, $\omega_i\tau_i \ll 1$ plasma jets \cite{hsu12pop} because the initially strong
magnetic field ($B^2/\mu_0 \sim \rho v^2$) at the gun nozzle decays by $1/e$
every few microseconds \cite{merritt14}
due to the high density ($> 10^{16}$~cm$^{-3}$) and low $T_e \approx
1.5$~eV\@.  Intuitively, these observations suggest that the merging 
of gun-formed plasmas using an intermediate value of
$\lambda_{\rm gun}$, i.e., $\lambda_{\rm spheromak} < \lambda_{\rm gun} < \infty$,
could potentially lead to a merged plasma with $\beta > 1$, $\omega_i\tau_i \gtrsim 1$.
In an eventual integrated experiment with a plasma liner compressing a target, 
it is envisioned that the relative initiation times and speeds of the target- and liner-formation
jets can be chosen such that the incoming liner is able to engage the stagnated
target before the target can expand very much.  Further studies are needed to
determine whether this is feasible and how much target expansion prior to liner engagement
is tolerable.

The above suggests a research path (1)~employing 3D single- and two-fluid MHD
simulations to explore the $\lambda_{\rm gun}$ parameter space in order to
identify whether the formation of $\beta> 1$, $\omega_i\tau_i\gtrsim 1$ plasmas is possible
via merging multiple gun-formed plasmas, and (2)~performing experiments merging
two gun-formed plasmas over a range of $\lambda_{\rm gun}$ values, guided by the
simulations, and characterizing $\beta$, $\omega_i\tau_i$, and $\ell$ via diagnostic measurements.
The initial simulations could be performed, e.g., using the LA-COMPASS (3D MHD) \cite{li03},
the USim (multi-fluid MHD) \cite{beckwith15}, and/or the FLASH (3D rad-MHD) \cite{fryxell00}
codes.  The initial two-plasma-merging experiments are
being planned for execution at the Wisconsin Plasma Physics Laboratory (WiPPL)
user facility \cite{forest15}.
Assuming success in this initial research phase, the next step
would be to add external coils
or permanent magnets to the existing PLX coaxial guns \cite{witherspoon17} in order
to apply an appropriate value of $\psi_{\rm gun}$, and to
form a $\beta>1$, $\omega_i\tau_i \gtrsim 1$ magnetized target plasma
by merging an array of 6--12 supersonic hydrogen
or deuterium plasma jets.  This could be performed at the PLX facility \cite{hsu15jpp,hsu18ieee},
where, with a higher number of merging jets, we could also study the feasibility of creating  
tangled fields with $\ell_0 \ll r_0$.

If it turns out to be impossible or overly difficult
to create the $\beta >1$, $\omega_i\tau_i \gtrsim 1$ conditions
by adjusting $\lambda_{\rm gun}$, as proposed above, an alternative plan
aims to independently apply a magnetic field to an unmagnetized
plasma that is first formed by merging $\beta \gg 1$, $\lambda_{\rm gun}=\infty$ plasma jets
\cite{hsu12ieee}.
The magnetic field could potentially be seeded by laser-generated beat-wave current drive
\cite{welch12,welch14}, whereby two lasers with slightly offset frequencies would
create a beat wave with a frequency of order the thermal electron plasma frequency.
This process would resonantly accelerate thermal electrons,
driving electrical current and generating a magnetic field, as has been
shown in 2D electromagnetic particle-in-cell simulations \cite{welch12,welch14}
in PJMIF-relevant regimes.  This approach may be better suited to create
a $\beta>1$ target with closed field lines, e.g., driving electrical current down one axis of a
spherical target to create an azimuthal field.  Much research
is needed to establish the feasibility of this alternate plan. Proof-of-concept experiments
to demonstrate small-scale, beat-wave magnetization (using 1.064- and 1.053-$\mu$m lasers)
of an initially unmagnetized dense plasma ($\sim 10^{18}$~cm$^{-3}$) are
underway \cite{yates17} on the Janus laser at the Jupiter Laser Facility at 
Lawrence Livermore National Laboratory.  The use of charged-particle beams rather than
laser-generated beat waves to magnetize a target plasma
should also be investigated, e.g., building on prior
work that demonstrated FRC formation using an electron beam without a pre-applied magnetic field
\cite{sethian78prl}.

\section{Feasibility of a Near-Term, Target-Heating Experiment}
\label{sec:experiment}

In this section, assuming success with liner and target formation based on the
the existing generation of plasma guns \cite{witherspoon17,hsu18ieee},
we evaluate whether a near-term, proof-of-concept,
target-compression experiment is capable of demonstrating target heating as
an important milestone for PJMIF (using the same coaxial plasma guns).  This near-term target-heating
experiment would use a subscale liner \cite{hsu18jfe} to compress a subscale target.

\begin{figure}[!b]
\centerline{\includegraphics[width=2.3truein]{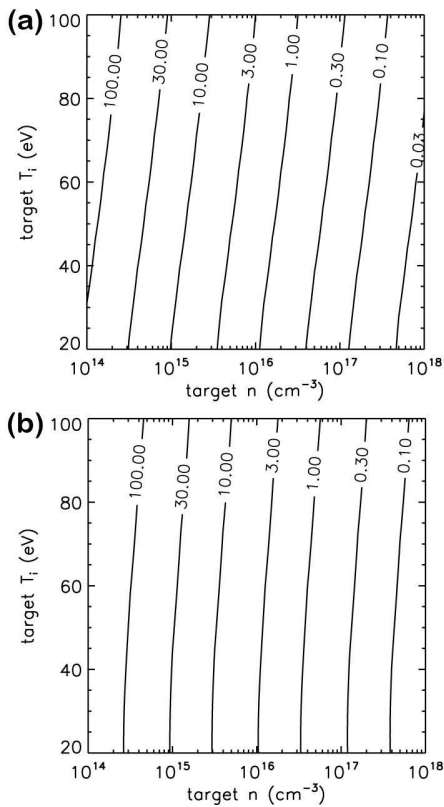}}
\caption{\label{fig:slowing_distance}Contours of penetration distance (cm) of liner argon ions
into the deuterium target plasma vs.\ target $T_i$ and $n$ for liner implosion
speed (a)~$v_0=60$~km/s and (b)~80~km/s.}
\end{figure}

A requirement that constrains the target parameters and liner implosion speed is
that the liner must act like a good piston, i.e., its penetration into the target must be small
compared to the target radius.  The penetration is set by the ion--ion frictional slowing distance of
a liner ion (e.g., Ar) into the deuterium target plasma (the frictional
slowing distance of Ar on electrons is much larger in the regimes of interest). 
Figure~\ref{fig:slowing_distance}
shows contours of Ar--D slowing distance as a function of target $T_i$ and $n$ for
subscale-relevant $v_0=60$~km/s and 80~km/s, respectively,
where the slowing distance is $l_s=v_0/(4\nu_s^{\rm Ar|D})$ \cite{messer13} and 
$\nu_s^{\rm Ar|D}$ is the slowing rate of argon ions on deuterium plasma (mean-charge
state $Z=1$ is assumed for both liner and target),
as given in the NRL Formulary (p.~31, 2016 edition) \cite{nrl-formulary}.  Anticipating that the
subscale target will have a diameter of $\sim 10$~cm, it is reasonable to require
that $l_s \lesssim 1$~cm for the liner to act as an effective piston.  
The piston requirement is easier to satisfy for a fusion-scale target, which has much higher
$n\sim10^{18}$~cm$^{-3}$.

Assuming that the subscale liner will have $v_0\approx 60$~km/s, 
Fig.~\ref{fig:slowing_distance}(a) shows that a target $n\gtrsim 10^{16}$ is required
for $l_s \lesssim 1$~cm.   To determine whether near-term,
pre-compression target parameters with $n\sim 10^{16}$~cm$^{-3}$ can
be realized, we consider the problem of merging 6--12 deuterium plasma jets
(consistent with the achieved plasma-jet parameters \cite{witherspoon17,hsu18ieee})
to form a ``target liner,'' which (upon stagnation) results in the subscale target.
This is similar to formation of the ``compression liner'' that will compress the target.  The main 
differences are using fewer
jets (6--12 rather than 36--60) and hydrogen or deuterium (rather than argon, krypton, or xenon)
for the jet species.

Finally, we repeat the same thermal-transport and magnetic-dissipation
analyses presented in the ``Fusion-Scale PJMIF Target'' section for
the subscale target plasma.  This sets
requirements on the subscale liner to meet the conditions for target heating, while 
simultaneously satisfying the requirement for the liner to act like a good piston.
Finally, we conduct
1D simulations [that also include radiation and equation-of-state (EOS) effects]
of the liner compressing the target to verify that target heating occurs.

\subsection{Initial Conditions of the Target-Formation Liner}
\label{sec:target_liner}

To conduct a 1D simulation of an imploding target liner to form
a subscale target plasma, we must first determine its initial
conditions, based on the merging of 6--12 near-term, achievable plasma jets.
Initial conditions of the target liner consist of its inner radius $r_{TL0}$, thickness $\Delta_{TL0}$,
velocity $v_{TL0}$, temperature $T_{TL0}$ (where $T_{e0}=T_{i0}$ is assumed), and
ion number density $n_{TL0}(r)$, where the subscript ``TL'' refers to target liner.
All quantities except $r_{TL0}$ are approximately known or derivable from
the achievable plasma-jet parameters.  The quantity $r_{TL0}$ is determined from (and
equivalent to) the merging radius $r_m$ of the target-formation jets, according to \cite{cassibry13}
\begin{equation}
r_{m,max} = \frac{r_{j0}[M_j (\gamma - 1)/2 + 1] + r_w}{1+(2/N^{1/2})[M_j (\gamma -1)/2+1 ]},
\label{eq:rm_max}
\end{equation}
where $r_{j0}$ is the initial jet radius at the chamber wall (where jets are launched),
$M_j$ the initial jet Mach number, $r_w$ the chamber-wall radius, and $N$ the number of jets.
Equation~(\ref{eq:rm_max}) assumes that the jet expands both radially and axially at
the speed $2C_s/(\gamma - 1)$.  If the jets expand instead at the slower speed $C_s$, 
which appears to be a better match to experiments \cite{hsu12pop}, then 
\begin{equation}
r_{m,min} = \frac{r_{j0}(M_j + 1) + r_w}{1+2(M_j+1)/N^{1/2}}.
\label{eq:rm_min}
\end{equation}
Because the target-formation jets have $\beta > 1$, this
means that $C_s > V_A$, where $V_A$ is the Alfv\'en speed, and thus we do not additionally
consider explicitly jet expansion at speed $V_A$.

\begin{figure}[!b]
\centerline{\includegraphics[width=2.3truein]{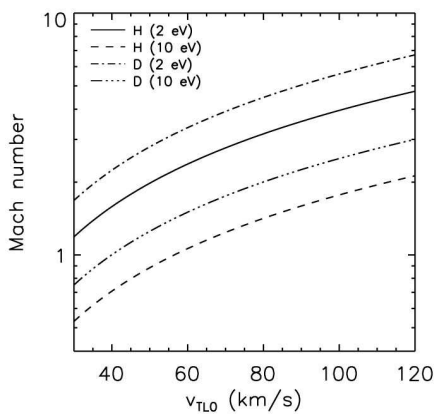}}
\caption{\label{fig:mach}Mach number $M\equiv v_{TL0}/C_s$ vs.\ jet speed $v_{TL0}$
for four cases (indicated in the legend),
where jet ion sound speed $C_s=[\gamma k(T_e+T_i)/m]^{1/2}$, and $\gamma=5/3$ and $T_e=T_i$ 
are assumed.}
\end{figure}

\begin{figure*}[!t]
\centerline{\includegraphics[width=5.7truein]{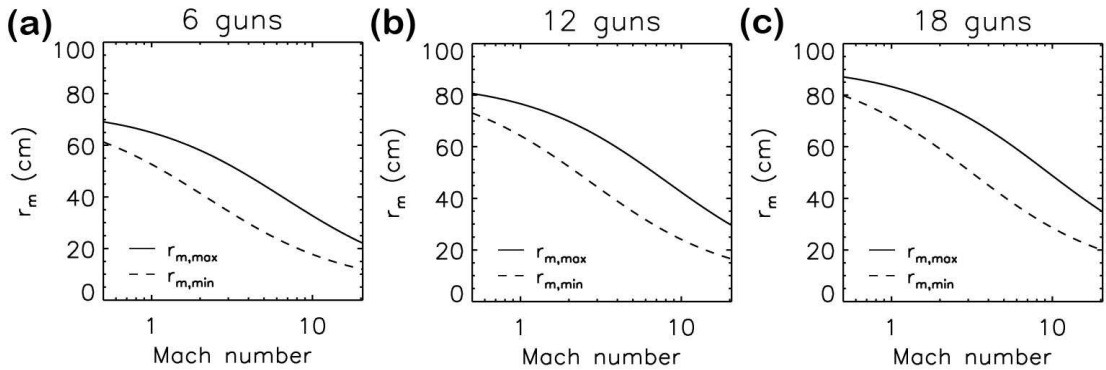}}
\caption{\label{fig:rm} Target-liner merging radii $r_{m,max}$ and $r_{m,min}$
vs.\ $M$ for (a)~$N=6$, (b)~$N=12$, and (c)~$N=18$ jets,
from Eqs.~(\ref{eq:rm_max}) and (\ref{eq:rm_min}), respectively,
assuming $\gamma=5/3$, $r_{j0}=4.25$~cm, and $r_w = 130$~cm.}
\end{figure*}

To evaluate $r_m$, we first need to know $M_j$.  Figure~\ref{fig:mach} shows the sonic Mach 
number $M$ vs.\ speed $v_{TL0}$ of target-formation
jets consisting of hydrogen or deuterium plasma at the
bounding cases of 2 and 10~eV\@.  For $v_{TL0} = 50$--100~km/s,
Fig.~\ref{fig:mach} reveals that $M$ spans the range $\approx 0.9$--6 for 
hydrogen at 10 eV to deuterium at 2 eV\@.  For this range of $M$,
evaluation of Eqs.~(\ref{eq:rm_max}) and (\ref{eq:rm_min}) for $N=6$, 12, and 18 jets, 
assuming $\gamma=5/3$, $r_{j0}=4.25$~cm (corresponding to the existing PLX guns/jets),
and $r_w=130$~cm (corresponding to the PLX vacuum chamber),
tells us the range of $r_m$ to be expected, as shown in Fig.~\ref{fig:rm}.

\begin{figure}[!t]
\centerline{\includegraphics[width=2.2truein]{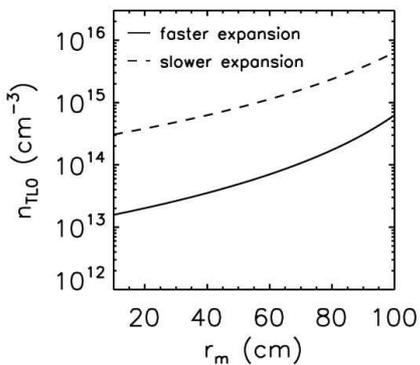}}
\caption{\label{fig:nTL0}Target-liner (deuterium) ion density $n_{TL0}(r_m)$ vs.\ $r_m$, assuming
$v_{TL0}=60$~km/s, $T_i=T_e=3$~eV, $\gamma = 5/3$, $n=10^{17}$~cm$^{-3}$,
$r_{j0}=4.25$~cm, initial jet length $L_{j0}=5$~cm, and $r_w=130$~cm.
Faster and slower expansion in the legend refers to expansion at $2C_s/(\gamma-1)$
and $C_s$, respectively.}
\end{figure}

We choose $r_m=55$~cm, corresponding to $N=12$ and $M\approx 3$, as a representative
case.  The value $n_{TL0}(r_m)$ is determined from the amount of jet-volume $V$ expansion as the
jet travels from $r_w$ to $r_m$, according to $n_{TL0}(r_m)= n_{j0} V(r_w)/V(r_m)$, shown in 
Fig.~\ref{fig:nTL0}.
We choose $n_{TL0}(r_m)=3\times 10^{14}$~cm$^{-3}$, corresponding to the slower-expansion
case for deuterium at $r_m=55$~cm (see Fig.~\ref{fig:nTL0}).
Because individual jets are coming together over $4\pi$ solid angle, we impose
a 1D liner-density
profile $n_{TL0}(r) = n_{TL0}(r_m) Nr_j^2/(4r^2)$ (for $r\ge r_m$), where $r_j\approx 31.6$~cm
and length $L_j=30$~cm
at $r_m=55$~cm (for the slower-expansion case) and $4\pi r_m^2 \equiv N \pi r_j^2 \rightarrow r_m^2/r_j^2 \equiv N/4$.
Thus, the mass of the
1D target liner is equal to the total mass of the $N=12$ jets, i.e., mass $=N\pi r_j^2(r_m)
L_j(r_m)n_{TL0}(r_m)m_D$.
Table~\ref{table:target_liner} summarizes all the parameters (based on expected target-formation jet parameters) that comprise the initial target-liner conditions to be used in a 1D implosion simulation
to determine the pre-compression, subscale deuterium-target parameters.
\begin{table}[!t]
\caption{\label{table:target_liner}Representative target-liner initial conditions 
used in 1D implosion calculations of the target liner for forming a near-term, subscale
deuterium plasma target.  We assume that $N=12$ plasma jets merge to form the
target liner.}
\begin{center}
\begin{tabular}{lc}
\hline\noalign{\smallskip}
parameter & value\\
\noalign{\smallskip}\hline\noalign{\smallskip}
$r_{TL0}=r_m$ & 55~cm\\
$\Delta_{TL0}=L_j(r_m) $ & 30~cm\\
$v_{TL0}$ & 60~km/s\\
$T_{TL0}$ & 3~eV\\
$n_{TL0}(r_m)$ & $3\times 10^{14}$~cm$^{-3}$\\
$n_{TL0}(r) $ & $n_{TL0}(r_m) Nr_j^2/(4r^2)$\\
\hline
mass & 1.1~mg\\
kinetic energy & 2.0~kJ\\
\hline
\end{tabular}
\end{center}
\end{table}

\subsection{Simulation of Target-Liner Implosion to Form a Subscale Target Plasma}

We use the initial conditions
given in Table~\ref{table:target_liner} to simulate the implosion
and stagnation of a target liner, which forms the plasma target, to determine
the plasma parameters of the pre-compression target.  We use the 1D
radiation-hydrodynamics code HELIOS \cite{macfarlane06}, which
has detailed EOS modeling capabilities. HELIOS cannot
model magnetic fields in spherical geometry, and thus we used a multiplier of 0.5 to the code's Spitzer thermal-conductivity model as a way to phenomenologically 
capture the effects of magnetized thermal transport in the target-liner implosion and stagnation.
The liner is modeled using 300 computational
zones (initial average of 1~mm/zone) with automatic zone refinement, separate $T_e$ and
$T_i$ evolution (2$T$), both radiation and thermal transport (with a multiplier of 0.5),
and non-LTE (local thermodynamic equilibrium)
EOS and opacity tables generated using PROPACEOS \cite{macfarlane06}.  The ``vacuum'' region
initially at $r<r_{TL0}=55$~cm is modeled identically as the liner itself, also using
300 zones, but with the following differences:  (a)~$n=3\times 10^{11}$~cm$^{-3}$,
(b)~initial velocity $v(r) = -v_{TL0} (r/r_{TL0})^2$ that decreases from
$v(r=0)=0$ to $v(r=r_{TL0})=-v_{TL0}$, and (c)~thermal-conductivity multipler of 1.0 
rather than 0.5.  Modeling
the ``vacuum'' region in this manner mitigates the artificial effects of a strong reflected shock
arising from the fast-expanding, leading edge of the liner reaching the origin first.

\begin{figure}[!t]
\centerline{\includegraphics[width=2.3truein]{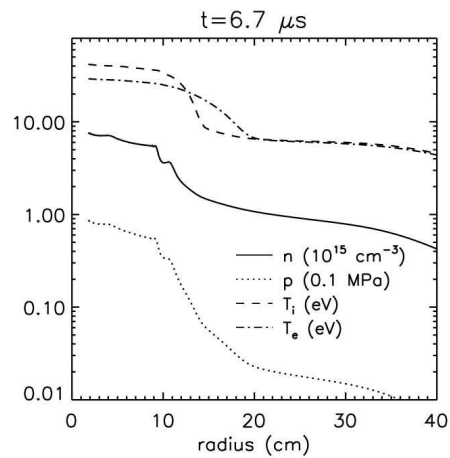}}
\caption{\label{fig:helios_profiles}Quantities in the legend vs.\ 
radius at $t=6.7$~$\mu$s (corresponding
to the time of peak thermal pressure $p$, where $t=0$ is when the leading
edge of the target-liner is at $r_m=55$~cm), from a HELIOS 1D
simulation (using the initial conditions of Table~\ref{table:target_liner}) of an imploding deuterium target 
forming a deuterium plasma target.  }
\end{figure}

HELIOS simulation results of the 1D radial profiles of several
plasma quantities at $t=6.7$~$\mu$s are shown in Fig.~\ref{fig:helios_profiles}, corresponding
to the time of peak thermal pressure.
Examining these profiles reveals that target-liner implosion results in a
deuterium plasma with peak $n \approx 7.6 \times 10^{15}$~cm$^{-3}$,
radius $a \approx 9.4$~cm as inferred from the HWHM of
$p(r)$, peak $T_i\approx 42$~eV, peak $T_e \approx 29$~eV, 
and lasting for $\approx 2.1$~$\mu$s as
inferred from the FWHM of $p(t)$ (not shown here).
The inferred subscale, pre-compression deuterium-target
parameters and other derived/chosen ones are summarized in Table~\ref{table:target}.

\begin{table}[!b]
\caption{\label{table:target} Summary of near-term, subscale, pre-compression, deuterium
target parameters (compare with fusion-scale target parameters of Table~\ref{table:PJMIF_target}).}
\begin{center}
\begin{tabular}{ll}
\hline\noalign{\smallskip}
parameter & value\\
\noalign{\smallskip}\hline\noalign{\smallskip}
$r_0$ & 9.4~cm\\
$n_0$ & $7.6 \times 10^{15}$~cm$^{-3}$\\
$T_{i0}$ & 42~eV\\
$T_{e0}$ & 29~eV\\
$p_0$ & 0.09~MPa\\
$B_0$ & 1.47~kG\\
$E_0$ & 451~J\\
\hline
$\beta$ & 10 \\
$\omega_i\tau_i$  & 0.74\\
$\omega_e\tau_e$ & 20\\
$\lambda_i$ & 0.47~cm\\
$\lambda_e$ & 0.18~cm\\ 
$\rho_i$ & 0.64~cm\\
$\rho_e$ & 0.009~cm\\
$\ell_0$ & 2.0~cm\\
$v_0$ & 60~km/s\\
\noalign{\smallskip}\hline
\end{tabular}
\end{center}
\end{table}
Referring to Fig.~\ref{fig:slowing_distance}(a), we see that the penetration of a 60-km/s imploding
argon compression liner into this subscale target is $\approx 2$~cm, which allows the
compression liner to act like a reasonably good piston.

\subsection{Properties of the Near-Term, Subscale Target}

\begin{figure}[!b]
\centerline{\includegraphics[width=2.3truein]{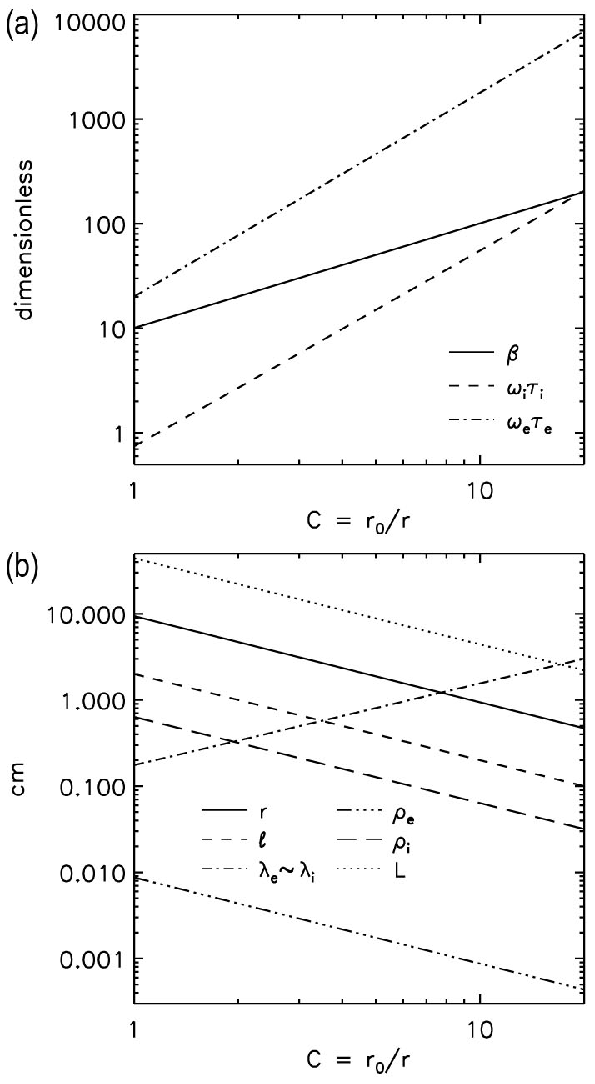}}
\caption{\label{fig:subscale_params}Quantities in the legend
vs.\ $C$ for the subscale target parameters in Table~\ref{table:target},
assuming adiabatic heating.}
\end{figure}

In this section, we repeat the analyses presented earlier
in the ``Fusion-Scale PJMIF Target'' section
to evaluate the thermal-transport and magnetic-dissipation properties of the near-term
pre-compression deuterium target of Table~\ref{table:target}, in order
to determine whether target heating via compression by a subscale plasma liner is possible.
Figure~\ref{fig:subscale_params} shows various dimensionless quantities vs.\ $C$ for
spherical adiabatic compression of this subscale deuterium-plasma target.

\subsubsection{Macro-stability of a Subscale Target}
\label{sec:stability}

As before with the fusion-scale target, we aim to create a subscale target that 
has pre-compression $\beta \sim 10$.  If we are successful, then MHD instabilities will likely be 
sidestepped, and the hydrodynamic disassembly time
($\sim 2.1$~$\mu$s for the hypothetical, subscale target in Table~\ref{table:target}) becomes a 
bottleneck in that the incoming liner must compress the target before it can disassemble
(or shortly thereafter).  It must also compress the target in a short-enough
time that overcomes the thermal loss rate and magnetic dissipation during the target
convergence through stagnation.  

\subsubsection{Thermal Transport in a Subscale Target}
\label{sec:thermal}

\begin{figure}[!t]
\centerline{\includegraphics[width=2.4truein]{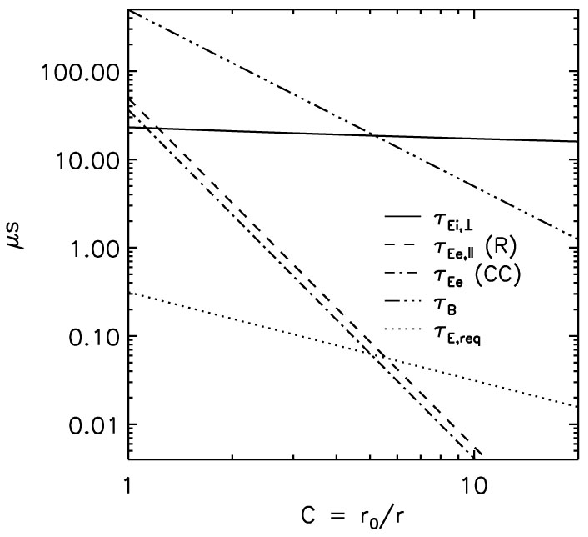}}
\caption{\label{fig:tau_vs_C_subscale} Quantities in the legend vs.\ $C$ for
the subscale target parameters in Table~\ref{table:target}, assuming adiabatic heating.}
\end{figure}

Using the initial target parameters in Table~\ref{table:target}, we calculate
$\tau_{Ei,\perp}$, $\tau_{Ee,\|} {\rm (R)}$, $\tau_{Ee} {\rm (CC)}$,
$\tau_B$, and $\tau_{E,req}$ vs. $C$,
as shown in Fig.~\ref{fig:tau_vs_C_subscale} (see the discussion in the
``Thermal Transport in a Fusion-Scale Target'' section).
For a subscale target with closed field lines, perpendicular transport is acceptable for
adiabatic heating to $C>10$.  For a subscale
target with tangled, open field lines, the electron thermal transport is acceptable for
adiabatic heating up to $C\approx 5$, which should allow for observation of target heating.
See the appendix for a more detailed treatment of the limits of target adiabatic heating.

\begin{figure}[!b]
\centerline{\includegraphics[width=2.4truein]{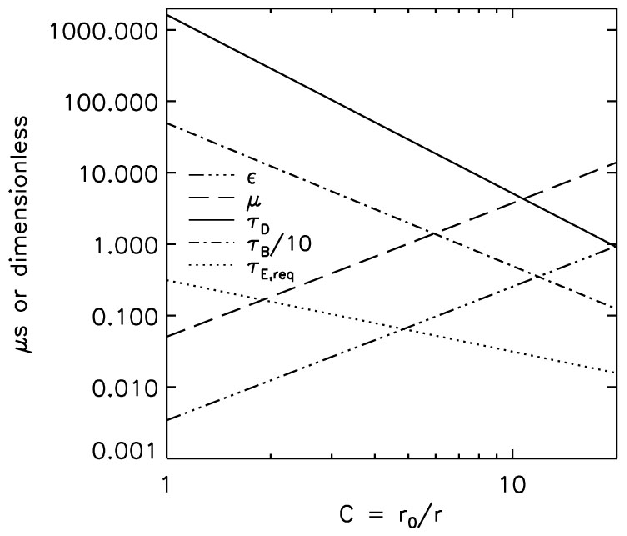}}
\caption{\label{fig:tauD_subscale}Quantities in the legend vs.\ $C$ for the subscale
target parameters in Table~\ref{table:target}, assuming adiabatic heating.}
\end{figure}

Figure~\ref{fig:tauD_subscale} shows the relevant quantities for evaluating
anomalous transport in a subscale target (with closed field lines) due to drift instabilities \cite{ryutov02pop}.
As seen in Fig.~\ref{fig:tauD_subscale},
the condition $\mu\ll 1$ is violated around $C\approx 5$ (see the
discussion in the ``Thermal Transport in a Fusion-Scale Target'' section).
Thus, it is possible that $\tau_B/10$ becomes the relevant perpendicular diffusion time for
$C\gtrsim 5$.  Nevertheless, $\tau_{E,req} \ll \tau_B/10$ up to
$C>10$, and thus drift-instability-induced anomalous transport is not expected
to play a substantial role.

Given this analysis, we conclude
that in a near-term subscale target-compression experiment (with the initial target conditions of
Table~\ref{table:target}),
adiabatic heating may theoretically be observed up to $C\approx 5$ (depending
on transport model used) for the case of tangled,
open field lines and to much higher $C$ for closed field lines,
both assuming that target compression is initiated
before it disassembles in a dwell time $\sim 2.1$~$\mu$s.
As before, ignoring radiative losses is justified, as $\tau_R/\tau_{E,req} \approx 1.6 \times 10^4$
at $C=5$.

\subsubsection{Magnetic-Energy Dissipation in a Subscale Target}
\label{sec:magnetic}

\begin{figure}[!b]
\centerline{\includegraphics[width=2.3truein]{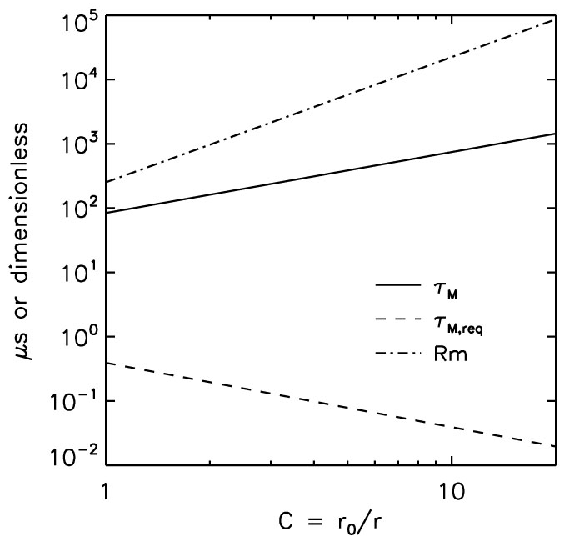}}
\caption{\label{fig:tauM_subscale}Quantities in the legend vs.\ $C$ for the subscale
target parameters in Table~\ref{table:target}, assuming adiabatic heating.}
\end{figure}

Figure~\ref{fig:tauM_subscale} shows $\tau_M$ and $\tau_{M,req}$ vs.\ $C$ 
for the subscale target (see the earlier discussion in the
``Magnetic-Energy Dissipation in a Fusion-Scale Target'' section),
showing that magnetic compression dominates over resistive decay.
Figure~\ref{fig:tauM_subscale} also shows that
$Rm\equiv \mu_0 \ell v_0/\eta \gg 10^2$ for the subscale target,
indicating that
the magnetic field is frozen into the plasma motion and that target compression
may lead to self-similar compression of any tangled field that is initially present,
as assumed in \cite{ryutov09}.  Also, using Eq.~(\ref{eq:tauN}) and the
parameters of Table~\ref{table:target}, we evaluate the
characteristic Nernst flux-loss time $\tau_N \approx 310$~$\mu$s, which
is $\gg \tau_{M,req}$ and therefore negligible.


Finally, we consider the condition, given by Eq.~(\ref{eq:u_lt_vti}), to avoid current-driven
anomalous resistivity \cite{davidson75} in the subscale target.
Using the parameters in Table~\ref{table:target}, we obtain
the requirement from Eq.~(\ref{eq:a_over_ell0}) that $\ell_0 > 1.2$~mm, which is satisfied by
the choice of $\ell_0=2.0$~cm.

\subsection{Subscale Plasma-Liner Compression of a Subscale Target}
\label{sec:target_compression}

In this subsection, we simulate 1D subscale-plasma-liner compression of the subscale target with
parameters given in Table~\ref{table:target}.
We follow a similar analysis as presented in the ``Initial Conditions of the Target-Formation
Liner'' section to determine $r_m$ and $n$ at $r_m$ for the 
compression liner, assuming achieved PLX-scale plasma-jet parameters.
Then, using the compression-liner parameters at $r_m$, we conduct a 1D HELIOS implosion
simulation to determine the compression-liner parameters at the moment it engages the
subscale target (with parameters given in Table~\ref{table:target}).  Finally, we conduct
another 1D HELIOS implosion simulation of the subscale liner compressing the subscale target
starting at the moment of liner/target engagement.

\subsubsection{Compression-Liner Initial Conditions}

\begin{figure}[!b]
\centerline{\includegraphics[width=2.3truein]{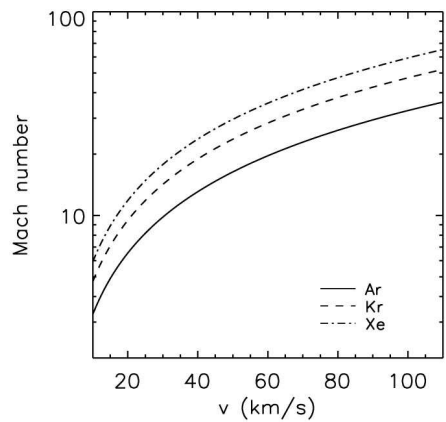}}
\caption{\label{fig:M_cl}Compression-liner Mach number
vs.\ velocity for three possible liner species,
assuming $T_e=T_i=1.5$~eV and $\gamma=1.3$.}
\end{figure}

\begin{figure*}[!htb]
\centerline{\includegraphics[width=5.5truein]{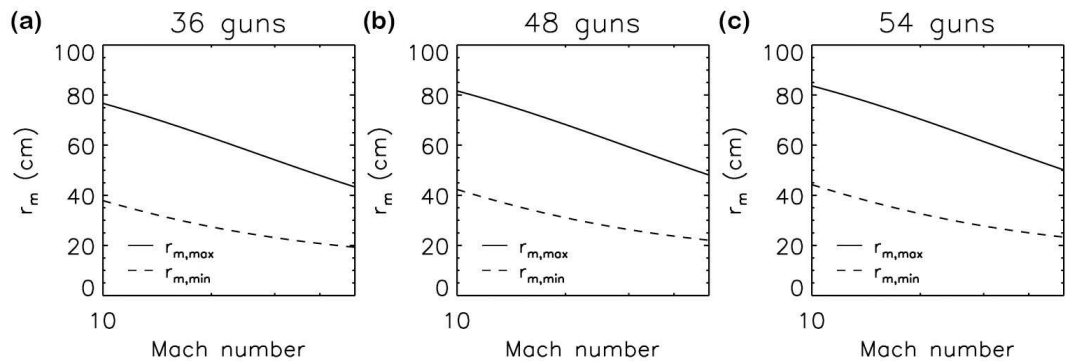}}
\caption{\label{fig:rm_cl}Compression-liner merging radii $r_{m,max}$ and $r_{m,min}$
vs.\ $M$ for (a)~36, (b)~48, and (c)~54 guns, from Eqs.~(\ref{eq:rm_max}) and (\ref{eq:rm_min}),
respectively, assuming $\gamma=1.3$ \cite{hsu18jfe}, $r_{j0}=4.25$~cm, and $r_w=130$~cm.}
\end{figure*}

Similar to the target-liner analysis, to determine the compression-liner merging radius $r_{m,CL}$,
we first evaluate $M$ of the liner as a function of candidate liner species and velocity, as shown in Fig.~\ref{fig:M_cl}, and $r_{m,CL}$ vs. $M$, as shown in Fig.~\ref{fig:rm_cl}.
By inspection of Fig.~\ref{fig:M_cl}, we see that $M\approx 20$ is a representative value for
argon (our preferred subscale liner species due to its lower cost compared to krypton
and xenon) and expected $v_0\approx 60$~km/s.  By inspection of Fig.~\ref{fig:rm_cl}, 
\begin{figure}[!tb]
\centerline{\includegraphics[width=2.4truein]{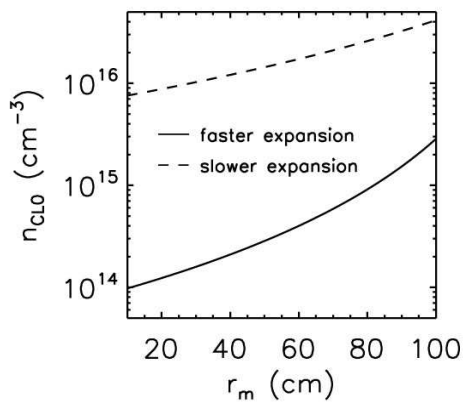}}
\caption{\label{fig:ncl0}Compression-liner (argon) density $n_{CL0}$ at $r_{m,CL}$ vs.\ $r_{m,CL}$,
assuming that the initial jet $v=60$~km/s, $T=1.5$~eV, $\gamma=1.3$ \cite{hsu18jfe},
$n=10^{17}$~cm$^{-3}$, $r_{j0}=4.25$~cm, length $L_{j0}=5$~cm, and $r_w=130$~cm.}
\end{figure}
we see that $r_{m,CL}\approx 40$~cm for $M\approx 20$ (for the slower-expansion case).
There is not a significant difference in $r_{m,CL}$ for
36 vs.\ 48 vs.\ 54 guns.  Figure~\ref{fig:ncl0} shows the density $n_{CL0}(r_m)$ vs.\ $r_{m,CL}$
for argon, and we choose $n_{CL0}\approx 4\times10^{15}$~cm$^{-3}$ for $r_{m,CL}=40$~cm.
Because $N=48$ jets are coming together, we impose a profile 
$n_{CL0}(r) = n_{CL0}(r_m) N r_j^2/(4r^2)$ (for $r\ge r_m$),
where $r_j\approx 11.5$~cm and $L_j\approx 10$~cm at $r_m=40$~cm,
and $4\pi r_m^2 \equiv N \pi r_j^2 \rightarrow r_m^2/r_j^2 \equiv N/4$.  Thus, the total mass
of the 1D compression liner is equal to the total mass of the $N=48$ jets, i.e., liner mass 
$=N \pi r_j^2(r_m) L_j(r_m) n_{CL0}(r_m) m_{Ar}$.
Table~\ref{table:compression_liner} summarizes
the initial conditions of the subscale compression liner at $r_{m,CL}$.

\begin{table}[!b]
\caption{\label{table:compression_liner}Representative argon, subscale, compression-liner initial conditions for compressing a subscale deuterium plasma target.  We assume that
$N=48$ plasma jets merge to form the subscale compression liner.}
\begin{center}
\begin{tabular}{lc}
\hline\noalign{\smallskip}
parameter & value\\
\noalign{\smallskip}\hline\noalign{\smallskip}
$r_{CL0}=r_m$ & 40~cm\\
$\Delta_{CL0}=L_j(r_m)$ & 10~cm\\
$v_{CL0}$ & 60~km/s\\
$T_{CL0}$ & 1.5~eV\\
$n_{CL0}(r_m)$ & $4 \times 10^{15}$~cm$^{-3}$\\
$n_{CL0}(r)$ & $n_{CL0}(r_m)Nr_j^2/4r^2$\\
\hline
mass (Ar) & 53.3~mg\\
kinetic energy (Ar) & 96~kJ\\
\hline
\end{tabular}
\end{center}
\end{table}

\subsubsection{Compression-Liner Parameters at the Moment of Target Engagement}

\begin{figure}[!b]
\centerline{\includegraphics[width=2.3truein]{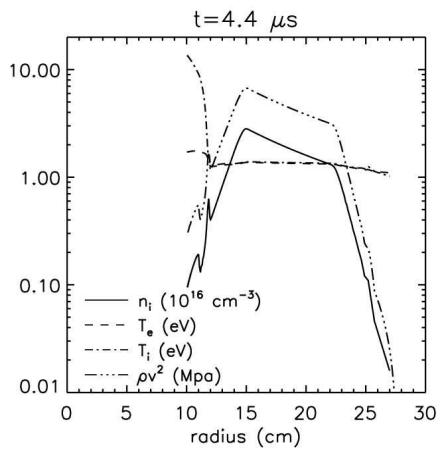}}
\caption{\label{fig:CL_engagement}Quantities in the legend
vs.\ radius at $t=4.4$~$\mu$s ($t=0$ corresponds to when the leading edge of the
liner is at $r_m=40$~cm) from a 1D HELIOS simulation of the imploding, subscale, argon
compression liner with initial conditions given in Table~\ref{table:compression_liner}.  Data
in the first 300 zones corresponding to the ``vacuum'' region have been set to zero.}
\end{figure}

To determine the compression-liner parameters at the moment of target engagement,
we perform a 1D HELIOS implosion simulation of the compression liner using the
initial conditions given in Table~\ref{table:compression_liner}.  In this simulation,
the argon liner is modeled using 300 computational zones 
(initial average of 0.3~mm/zone) with automatic zone refinement,
2$T$, radiation and thermal transport 
(with a multiplier of 1.0 for the Spitzer conductivity model),
and non-LTE EOS and opacity tables from PROPACEOS\@.  The ``vacuum'' region,
where $r<r_{CL0}$, is modeled identically as the compression liner, with the following
differences:  (1)~$n=1.5\times10^{12}$~cm$^{-3}$ and (2)~initial velocity 
$v(r)= -v_{CL0} (r/r_{CL0})^2$.
Figure~\ref{fig:CL_engagement} shows the simulation results of
the radial profiles of several plasma quantities
when the leading edge of the liner reaches $\approx 9.4$~cm, which is
when it should engage the pre-compression target in Table~\ref{table:target}.  

\subsubsection{Target Heating}

Finally, we use idealized versions (i.e., spatially uniform step functions)
of the subscale target and liner profiles
in Figs.~\ref{fig:helios_profiles} and \ref{fig:CL_engagement}, respectively,
to conduct 1D simulations of the subscale liner imploding the subscale target, in order
to verify the feasibility of compressional target heating in a near-term, subscale experiment.
The simulations include the effects of radiative losses and non-LTE EOS\@.
The idealized initial conditions of the liner engaging the target are given in Table~\ref{table:engagement}.  The mass of the idealized compression liner (52.6~mg)
agrees well with that of the initial conditions (53.3~mg) in Table~\ref{table:compression_liner}.
However, the mass of the idealized target (0.3~mg) is substantially lower than that of the
initial conditions (1.1~mg) of the target liner in Table~\ref{table:target_liner} because we
are not including the substantial radial ``wing'' in density beyond $r\approx 13$~cm (see Fig.~\ref{fig:helios_profiles}).
\begin{table}[!b]
\caption{\label{table:engagement}Idealized initial conditions for 1D simulations
of a subscale argon liner engaging a subscale deuterium target, based on the
target and liner profiles of Figs.~\ref{fig:helios_profiles} and \ref{fig:CL_engagement},
respectively.}
\begin{center}
\begin{tabular}{lcc}
\hline\noalign{\smallskip}
parameter & target (D) & liner (Ar)\\
 & ($r=0$--13~cm) & ($r=13$--25~cm)\\
\noalign{\smallskip}\hline\noalign{\smallskip}
$n$ (cm$^{-3}$) & $10^{16}$ & $1.4\times 10^{16}$\\
$T_i$ (eV) & 40 & 1.3 \\
$T_e$ (eV) & 25 & 1.3\\
$v$ (km/s) & 0 & -60\\
$p$ (MPa) & 0.1 & $6\times 10^{-3}$\\
$\rho v^2$ (MPa) & 0 & 3.4\\
mass (mg) & 0.3 & 52.6\\
\hline
\end{tabular}
\end{center}
\end{table}

\begin{figure}[!t]
\centerline{\includegraphics[width=2.4truein]{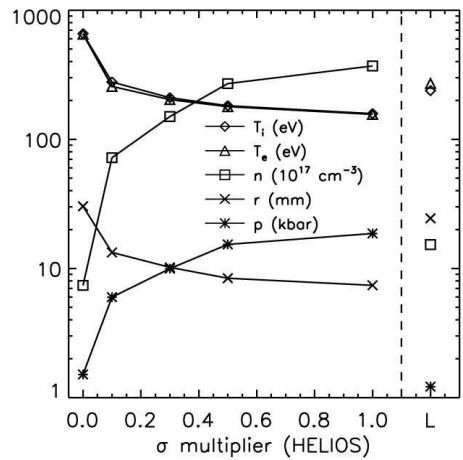}}
\caption{\label{fig:compression}HELIOS and Langendorf-model results of a subscale liner
compressing a subscale target using
the initial conditions of Table~\ref{table:engagement}; shown are target plasma quantities
indicated in the legend vs.\ the multiplier applied to the thermal conductivity $\sigma$,
at the time of peak fusion reactivity and at half the instantaneous fuel radius at that time.  The
values corresponding to the ``L'' column are the mean of the Langendorf results for
classical and Bohm transport.}
\end{figure}

We perform the calculations using both HELIOS and the Langendorf semi-analytic model
\cite{langendorf17pop}.  The latter includes more realistic
models for estimating the magnetized perpendicular
thermal transport for the case of closed field lines (but not the parallel transport for highly tangled, open
field lines), and assumes a magnetic field of 1.618~kG (corresponding
to $\beta= 10$). As mentioned earlier, because HELIOS does not model
magnetic fields in spherical geometry, we apply a multiplier to the code's
Spitzer thermal conductivity $\sigma$
as a way to model the reduced, perpendicular thermal transport in the target plasma.
The HELIOS and Langendorf-model
results, using the initial conditions given in Table~\ref{table:engagement}, are shown in
Fig.~\ref{fig:compression} and Table~\ref{table:compression}.
Results from the Langendorf model agree reasonably well with HELIOS results
using $0.1\sigma$, with both showing target compressional heating to over 200~eV\@.
Thus, we conclude that compressional
target heating in a near-term, subscale experiment is feasible.

The HELIOS case using $10^{-6}\sigma$ provides
an upper bound on target heating (including radiative losses) to $T=658$~eV
(at half the target radius) that compares well with the predicted adiabatic heating
to $T_{i0}C^2\approx 40\times 4.3^2 =740$~eV\@.
The case with $1.0\sigma$ shows that target heating may be observable
even without any magnetic insulation.
The peak temperatures achieved for all the HELIOS (except the $10^{-6}\sigma$ case) and Langendorf
results in Table~\ref{table:compression} are well below that predicted by adiabatic heating,
which was expected up to $C>10$ (for closed field lines), according to Fig.~\ref{fig:tau_vs_C_subscale}; 
further work is needed to identify the origin of the discrepancy.

\begin{table*}[!t]
\caption{\label{table:compression}HELIOS simulation results of the 1D 
implosion of the initial conditions of Table~\ref{table:engagement} for different
multipliers applied to the thermal conductivity $\sigma$.  For comparison,
Langendorf-model \cite{langendorf17pop} results
are also given for both perpendicular classical and Bohm transport models.  Results are at the time of
peak DD fusion reactivity at the spatial position equal to one-half of
the instantaneous target radius at that time.  The Langendorf model assumes a spatially uniform target.}
\begin{center}
\begin{tabular}{lccccccc}
\hline\noalign{\smallskip}
$\sigma$ & time of peak & target radius at & peak & $T_i$ & $T_e$ & $n$ & $p$\\
multiplier & reactivity ($\mu$s) & peak reactivity (cm) & C & (eV) & (eV) & ($10^{17}$~cm$^{-3}$) & 
(kbar)\\
\noalign{\smallskip}\hline\noalign{\smallskip}
$10^{-6}$ & 3.50 & 3.03 & 4.3 & 658 & 651 & 7.4 & 1.51\\
0.1 & 2.95 & 1.33 & 9.8 & 278 & 257 & 72 & 5.99\\
0.3 & 2.85 & 1.02 & 12.7 & 210 & 203 & 150 & 9.96\\
0.5 & 2.85 & 0.84 & 15.5 & 182 & 179 & 270 & 15.35\\
1.0 & 2.75 & 0.74 & 17.6 & 158 & 156 & 370 & 18.65\\
\hline
Langendorf classical & 2.89 & 2.56 & 5.1 & 262 & 323 & 13.2 & 1.23\\
Langendorf Bohm & 2.74 & 2.33 & 5.6 & 214 & 219 & 17.4 & 1.20\\
\hline
\end{tabular}
\end{center}
\end{table*}

\section{Conclusions and Future Work}

We describe the properties of
a novel, magnetized target plasma (with $\beta> 1$, $\omega_i\tau_i\gtrsim 1$, and 
possibly a tangled field with correlation length much smaller than the target radius)
that is well suited for compression by a high-implosion-speed, spherically imploding plasma liner.
We show that for the fusion-scale target
parameters of Table~\ref{table:PJMIF_target},
compressional adiabatic heating is possible to $C>10$ for
closed field lines.  For a target with highly, tangled open field lines, adiabatic heating to $C=10$ will be
challenging to achieve, and the upper limit on $C$ is sensitive to the details of the model being
used (see also the appendix).
We also show that magnetic dissipation, including Nernst effects, should be small, and that anomalous 
transport and resistivity arising from drift-induced
and current-driven instabilities, respectively, are not expected to be important.

Next, we describe a possible approaches to creating a $\beta> 1$, $\omega_i\tau_i\gtrsim 1$ 
plasma target, i.e., by merging multiple gun-formed plasmas using the parameter
$\lambda_{\rm gun}$ as a control knob.  Preparatory efforts are now underway to model and execute
an experiment to merge two gun-formed plasmas, with varying $\lambda_{\rm gun}$, 
at the WiPPL user facility to test this approach.  If this effort is successful, the next step
would be to merge 6--12 gun-formed plasmas at PLX to form a subscale $\beta> 1$ target
suitable for compression by a subscale liner.  Merging 6--12 guns on PLX will
allow us to explore the possibility of creating a tangled field with long
connection length in the target.  We also describe an alternative 
target-formation approach, i.e., form
an unmagnetized target plasma by merging multiple unmagnetized plasma jets, and then
independently magnetize the target via laser-generated beat-wave current drive.  Proof-of-concept
experiments to demonstrate the basic beat-wave magnetization physics are underway
using the Janus laser at the Jupiter Laser Facility at Lawrence Livermore National Laboratory.
Much research is needed to further assess both these target-formation approaches.

Finally, assuming we are able to form
a target with $\beta> 1$, $\omega_i\tau_i\gtrsim 1$, and possibly a tangled field,
we evaluate whether a near-term, proof-of-concept experiment
to demonstrate compressional
heating of such a target is feasible using the existing generation of coaxial plasma guns.  Using
achievable plasma-jet parameters, we estimate the achievable subscale target and liner parameters
(Table~\ref{table:engagement}),
and show theoretically that for a subscale target with closed field lines, adiabatic heating is theoretically possible to $C>10$. For a target with tangled, open field lines,
adiabatic heating is theoretically possible to to $C\approx 5$ (see also the appendix).
Using the parameters of Table~\ref{table:engagement} as initial conditions,
both HELIOS and the Langendorf semi-analytic model predict appreciable target heating
assuming closed field lines and perpendicular transport,
with good agreement in the peak $T_i\approx 270$~eV (at half the target radius) when HELIOS uses 
$0.1\sigma$.  

Issues requiring further detailed studies are many, but we mention just a few here as
priorities.  The highest priority is to perform modeling and experiments to determine
whether targets with $\beta>1$, $\omega_i\tau_i\gtrsim 1$, and either closed or open tangled fields
can indeed be formed by merging multiple plasma jets.
Optimization of target- and compression-liner speeds and their relative
firing times from the chamber wall are needed because the target-liner jets expand much
more quickly than the compression-liner jets.  The optimized parameters should then be used as
a guide for integrated 3D radiation-MHD simulations that include the firing and merging of both the 
target and liner jets, and their subsequent convergence to stagnation.  The 3D simulations
are also needed to assess the effects of non-uniformities at the liner/target interface, and how much
the non-uniformities degrade the target compression and heating
due to deceleration-phase Rayleigh-Taylor
instabilities (RTI)\@.  There may be some mitigating
factors for RTI in MIF compared to inertial confinement fusion (ICF) due to
there being (i)~a strong and possibly sheared magnetic field at the target/liner interface
and (ii)~much smaller convergence ratio and deceleration magnitude, both of which may
provide a larger window of tolerance for RTI in MIF compared to ICF; much
further research is needed on this important issue.
There is also a need for further detailed study of the thermal transport in a target with highly
tangled, open field lines.  The Ryutov and C\&C scalings used in this paper give substantially
different predictions (see also the appendix).
Finally, although this paper is largely focused on analysis of
the near-term, subscale target-heating
experiment, it is worth mentioning that the fusion-scale compression liner may have a dense, cold
``afterburner'' fuel layer \cite{thio99} at the leading edge.  This affects the inflight dynamics
of the compression liner as well as the subsequent liner/target engagement and target compression,
all of which require further study.

The results of this paper motivate and chart a near-term research path toward the subscale
demonstration of the formation of a novel magnetized
target with $\beta>1$ and $\omega_i\tau_i\gtrsim 1$, and the compressional heating of that target using
a spherically imploding plasma liner formed by merging hypersonic plasma jets.  The
existing generation of coaxial plasma guns, with some
minor modification (i.e., addition of bias flux), has the technological readiness
level to support this development effort, although much gun development is still needed
for a fusion-scale demonstration of PJMIF\@.

\appendix

\section{Target Adiabatic Heating}

Following \cite{davies17pop} but generalizing to spherical geometry, we estimate
the temperature $T_c$, at which which electron thermal losses equals compressional heating
for both the Ryutov \cite{ryutov09} and C\&C \cite{chandran98} transport scalings.
We use (based on the expressions given in \cite{nrl-formulary},
where $T$ is everywhere in eV and all other variables are in cgs units),
\begin{equation}
\nabla T_e = -T_e/r,
\end{equation}
the electron thermal conductivity
\begin{equation}
\kappa_{e,\|}= 3.2\frac{nkT_e}{\nu_e m_e},
\end{equation}
and the electron heat flux
\begin{equation}
q_e= - \frac{\kappa_{e,\|}}{f} \nabla(kT_e)\equiv \frac{K_0}{f}\frac{T_e^{7/2}}{r},
\end{equation}
where $k=1.60\times10^{-12}$~erg/eV, $f$ is an adjustment based on 
\cite{chandran98,albright01} (discussed further below), and
$K_0 = 3.09\times10^{9}/\ln\Lambda$~(in cgs units) is a slowly varying
function of $n_e$ and $T_e$\@.
For $n=n_e=n_i$ and $T=T_e=T_i$, and assuming that electron heat flux dominates over ion
heat flux, the instantaneous 0D target energy evolution is (integrating over the target
volume and using the divergence theorem)
\begin{equation}
\frac{\rm d}{{\rm d}t} \left(4\pi r^3 nkT\right) \approx -4\pi r^2 q_e - 4\pi r^2 pv = -4\pi r^2(q_e + pv).
\end{equation}
Using $nr^3=$~constant and $v={\rm d}r/{\rm d}t < 0$ gives
\begin{equation}
n_0r_0^3 v k\frac{{\rm d}T}{{\rm d}r} \approx -r^2 (q_e+pv)
=-\frac{K_0 r T^{7/2}}{f} - 2nkT v r^2,
\end{equation}
and, after re-arranging,
\begin{equation}
\frac{{\rm d}T}{{\rm d}r} \approx -\frac{K_0 r T^{7/2}}{fkn_0r_0^3v} - \frac{2T}{r}.
\label{eq:heating_requirement}
\end{equation}

For adiabatic heating, it is required that the magnitude of the
second term on the right hand side of Eq.~(\ref{eq:heating_requirement})
(compressional heating) dominates over the first
(electron heat loss),
which leads to the condition
\begin{equation}
T\ll \left(\frac{2fk n_0 r_0 |v| C^2}{K_0}\right)^{2/5} \equiv T_c,
\label{eq:Tc}
\end{equation}
where $T_c$
is the temperature (at a given $C$) at which compression heating balances electron thermal losses.
Based on \cite{chandran98,albright01},
we assume $f=3\ln(\ell/\rho_e)$ (Eq.~(7) of \cite{chandran98}),
which is independent of $C$ for adiabatic scaling,
when $\lambda_e < \ell$, and we assume $f=15\ln(\ell/\rho_e)$ when $\lambda_e \ge \ell$
and transport reduction due to mirror trapping becomes applicable.  The
factor of 5 difference is based on Fig.~7 of \cite{albright01}.

\begin{figure}[!t]
\centerline{\includegraphics[width=2.3truein]{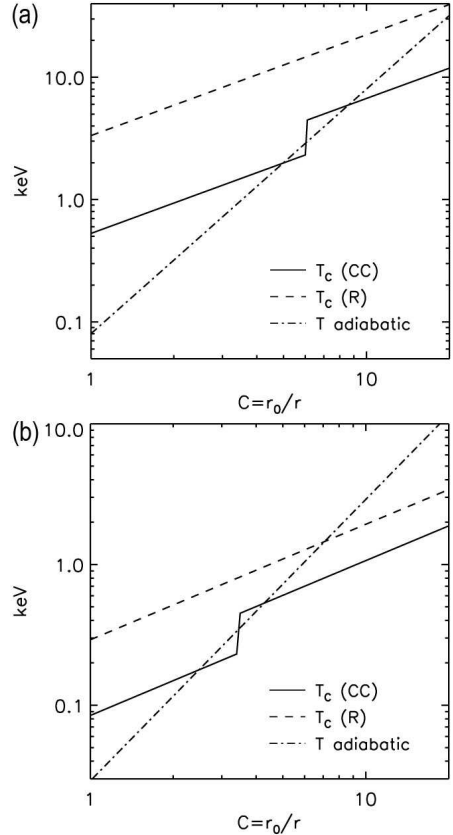}}
\caption{\label{fig:Tc}Temperature $T_c$ [Eq.~(\ref{eq:Tc})] vs.\ $C$ for (a)~the fusion-scale
target of Table~\ref{table:PJMIF_target} and (b)~the subscale target of Table~\ref{table:target},
for both the C\&C \cite{chandran98} and Ryutov \cite{ryutov09} transport scalings.  The
jogs in $T_c {\rm (CC)}$ is due to a factor of 5 reduction in thermal diffusivity due to 
mirror trapping when $\lambda_e \ge \ell$ beyond a threshold $C$\@.}
\end{figure}

Figures~\ref{fig:Tc}(a) and \ref{fig:Tc}(b) show $T_c$ vs.\ $C$
(for both the Ryutov and C\&C transport scalings) for the fusion-scale (Table~\ref{table:PJMIF_target})
and subscale (Table~\ref{table:target}) targets, respectively.  Equation~(\ref{eq:Tc})
defines $T_c$ for the C\&C scaling.  For the Ryutov scaling, the heat flux is multiplied
by a factor $(\ell/r)^2\ll 1$ by adjusting $f\rightarrow f(r/\ell)^2$; this
is due to the effect of the very long connection length of the parallel electron transport.
Including the benefit of mirror trapping in the C\&C scaling, 
the peak $C$ for adiabatic heating is slightly more pessimistic than that suggested by
the simpler analyses in the main text
underlying Figs.~\ref{fig:PJMIF_tauE_vs_C} and \ref{fig:tau_vs_C_subscale}.
The substantial difference between the Ryutov and C\&C scalings motivate further studies
of the thermal transport in targets with highly tangled, open field lines.

\begin{acknowledgements}
We thank Y. C. F. Thio, I. Golovkin, X.-Z. Tang, and D. Ryutov for discussions and advice,
and one of the anonymous referees for pointing out the possible relevance of the
more pessimistic Rechester \& Rosenbluth transport scaling in a stochastic magnetic field (compared
to the Ryutov scaling).
\end{acknowledgements}

\bibliographystyle{spphys}       

\end{document}